\documentclass[final,1p,times]{elsarticle} 
\usepackage{graphicx} 
\usepackage{amssymb} 
\usepackage{amsthm} 
\usepackage{lineno} 

\newcommand \ra  {\rightarrow}

\newcommand \sg {\sigma}

\newcommand \x {\cdot}

\newcommand \bvec{\left( \begin{array}{c} }
\newcommand \evec{\end{array} \right)}

\newcommand \bea{\begin{eqnarray} }
\newcommand \eea{\end{eqnarray} }

\newcommand \be {\begin{equation}}
\newcommand \ee {\end{equation}}

\journal{Nuclear Physics A}

\begin{document}

\begin{frontmatter}



\title{Jet-medium interactions at Quark Matter 09}


\author{Abhijit Majumder}

\address{Department of Physics, The Ohio State University, Columbus, Ohio, USA.}

\begin{abstract}
The suppression of the yield of high transverse momentum $p_{T}$ hadrons in heavy-ion collisions, 
referred to as ``jet-quenching'', has now developed into a comprehensive science. 
Jets are now used as probes of a variety of properties of the dense medium through which 
they propagate. Major theoretical improvements include jet modification in a 3-D fluid dynamical medium, 
the first set of in-medium Monte-Carlo implementations, 
an understanding of multi-hadron observables and energy flow within perturbative QCD, 
along with improvements in the AdS/CFT description of energy loss. On the experimental side, high statistics 
data are allowing for the first discriminatory test of various theoretical models and approximations 
while the new measurements of full jet reconstruction 
pose a challenge to theory.  
\end{abstract}


\end{frontmatter}



\section{Introduction}
\label{intro}

At Quark Matter 2001, the first data from the Relativistic Heavy-Ion Collider (RHIC) were presented~\cite{qm2001}: 
among others, it included the first hint of the suppression of the yield of $\pi^{0}$ at $p_{T} \geq $ 3 GeV (considered 
high $p_{T}$ at the time) in heavy-ion collisions compared to binary scaled $p$-$\bar{p}$. 
While strongly debated at the 
time, the modification of hard jets in dense matter (jet-quenching) has not only been established but 
has continued to develop as a tool to probe the dense matter produced in heavy-ion collisions. 
In these proceedings, we 
highlight the major advances reported at this meeting. 
 
This is by no means a complete or even-handed summary of all the advances in jet-medium interactions presented at this meeting. 
Many presentations had to be mentioned briefly or not at all due to space restrictions. 
Except for the special cases where only one exists, theoretical 
results and corresponding experimental data will be discussed simultaneously. 

\section{The evolving landscape in theory and experiment}

The modification of hard jets in dense matter is widely considered as a reliable probe of the medium 
produced in a heavy-ion collision. The weakening of the strong coupling constant with the scale of the 
hard collisions and the factorization properties of QCD allow for the use of a simple factorized form to 
calculate the inclusive cross section to produce a leading hadron with transverse momentum $p_{T}$ and rapidity $y$:
\bea
\frac{d^2 \sg^h}{dy d^2 p_T} = \frac{1}{\pi} \int dx_a \int d x_b \, G^A_a(x_a) G^B_b(x_b) 
\frac{d \sg_{ab \ra cX} }{d \hat{t}} \frac{\tilde{D}_c^h(z)}{z}.
\eea
In the equation above, $G^{A,B}(x)$ are the nuclear parton distribution functions to find collinear partons in the 
incoming nuclei, $d \sigma/d\hat{t}$ is the hard partonic cross section (with Mandelstam variable~$\hat{t}$) 
and $\tilde{D}^{h}(z)$  is the 
medium modified fragmentation function to produce a hadron from the hard parton.  The $\tilde{D}$ 
contains both the non-perturbative vacuum part and the modification 
to the jet parton distribution due to propagation through the medium.

The standard pQCD-based approaches invoke the presence of the hard scale and require that the hard partons 
from the jet couple weakly with the medium, allowing for the use of pQCD to describe these interactions. Related 
advances reported at this 
meeting included a computation of the energy momentum deposition, 
photon triggered away side yields, back-to-back correlations at NLO, a first attempt to understand jet 
shape observables and novel approaches in the budding Monte-Carlo implementations.


In recent efforts, spurred by the difficulty of pQCD-based approaches to accommodate heavy flavor suppression and conical emission, 
an entirely new approach has been invoked which assumes that the hard jet is strongly coupled with the medium.
These calculations use the AdS/CFT correspondence to compute both the drag experienced by a heavy quark in a strongly 
coupled medium and the ensuing wake which leads to conical flow. Major developments on these two fronts included the extension 
to light flavors and possible observation of Bragg peaks.

On the experimental side, the stream of high statistics data at high $p_{T}$ has continued at this meeting. 
There now exist a wide variety of data on all possible single and triggered yields, binned in terms of 
centrality, $p_{T}$, flavor and angle. Data on photon triggered away side yields, considered as the gold plated channel 
in jet quenching, were also presented by both STAR and PHENIX. New corrections to the open heavy flavor 
yields were reported by PHENIX along with the first attempt at an away side correlation with heavy flavors.

Perhaps the most exciting measurement reported at this meeting, also presented by both collaborations, 
was the possibility of full jet reconstruction in heavy-ion collisions. This also included jet $R_{AA}$
measurements by STAR and the first hint of intra-jet broadening. In the remainder we will present 
short discussions on the above mentioned topics.

\section{Success of the pQCD approach: The $R_{AA}, I_{AA}$ and $\hat{q}$ story}

The primary observable in the study of jet modification has been the suppression of the yield of single inclusive hadrons at 
high $p_{T}$ in a heavy-ion collision compared to a binary scaled $p$-$p$ collision. Central collisions demonstrate a factor 
of 5 suppression up to a $p_{T} \simeq 20$ GeV (measured in $\pi^{0}$) and have now been measured for all flavors of hadrons, i.e., $p (\bar{p})$, kaons, $\phi$'s, $\omega$'s and even for open heavy flavor etc~\cite{vale_putschke}. 
The slight differences between the various flavors are now eventually being understood as 
possibly due to jet conversions in a medium~\cite{fries}. 

\begin{figure}[!htb]
\hspace{0.5cm}
\resizebox{1.75in}{2in}{\includegraphics[2in,0.5in][8in,6.5in]{RAA_centrality.eps}} 
\hspace{0.5in}
\resizebox{2.in}{2.75in}{\includegraphics[0.in,0.2in][5.7in,6.0in]{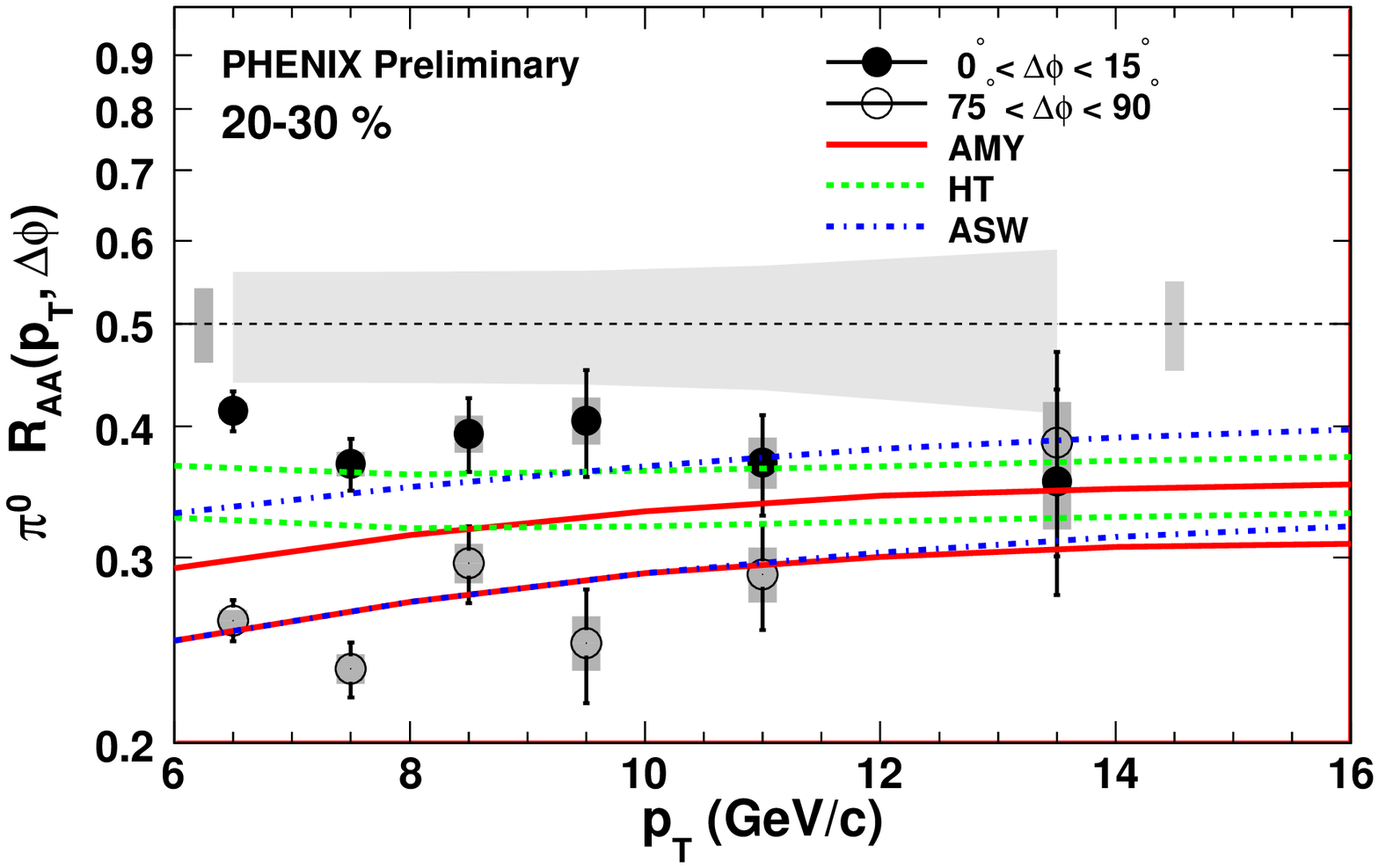}} 
  \caption{(Left plot) Calculations for the $R_{AA}$ vs. $p_{T}$  for the most central and semi-central collisions 
from the AMY, HT and ASW schemes (see text for details), in comparison with PHENIX data. (Right plot) Predictions from the same three approaches for the $R_{AA}$ in and out of plane v.s. $p_{T}$ compared to PHENIX data.}
    \label{fig1}
\end{figure}

The major component of the suppression for high $p_{T}$ $\pi^{0}$'s and its centrality dependence is now explained by 
all pQCD-based approaches as primarily due to radiative energy loss (see Ref.~\cite{jeon} for the most current 
review). In spite of the somewhat different 
approximations made in each of these calculations~\cite{Majumder:2007iu}, a rigorous study where three different 
formalisms were reduced to one parameter calculations and were forced to use an identical dynamical medium yielded 
the same $R_{AA}$ as a function of $p_{T}$ in both central and semi-central collisions~\cite{Bass:2008rv} (left plot of Fig.~\ref{fig1}).  
The schemes presented include one derived from the finite temperature field theory calculations of 
Arnold, Moore and Yaffe (denoted as AMY)~\cite{AMY}, one based on higher twist expansions of hard cross sections 
(denoted as HT)~\cite{HT}, and an eikonal formalism developed by Armesto, Salgado and Wiedemann 
(denoted as ASW)~\cite{ASW}.

Once the single parameter is set by comparison with one data point at one centrality, 
there remain small differences between the predictions from the different schemes. Restricting attention to 
only single inclusive observables, the primary difference is in the $R_{AA}$ as a function of the angle $\phi$ with the reaction plane~\cite{Majumder:2006we}. 
Experimental data for the $R_{AA} \, vs. \, \phi $ at high $p_{T}$ were compared to these calculations for the 
first time at this meeting allowing for the first discriminatory test of the models~\cite{Wei:2009mj} (right plot of Fig.~\ref{fig1}).

\begin{figure}[!htb]
\hspace{0.5cm}
\resizebox{2in}{2in}{\includegraphics[1in,0.5in][7.5in,7in]{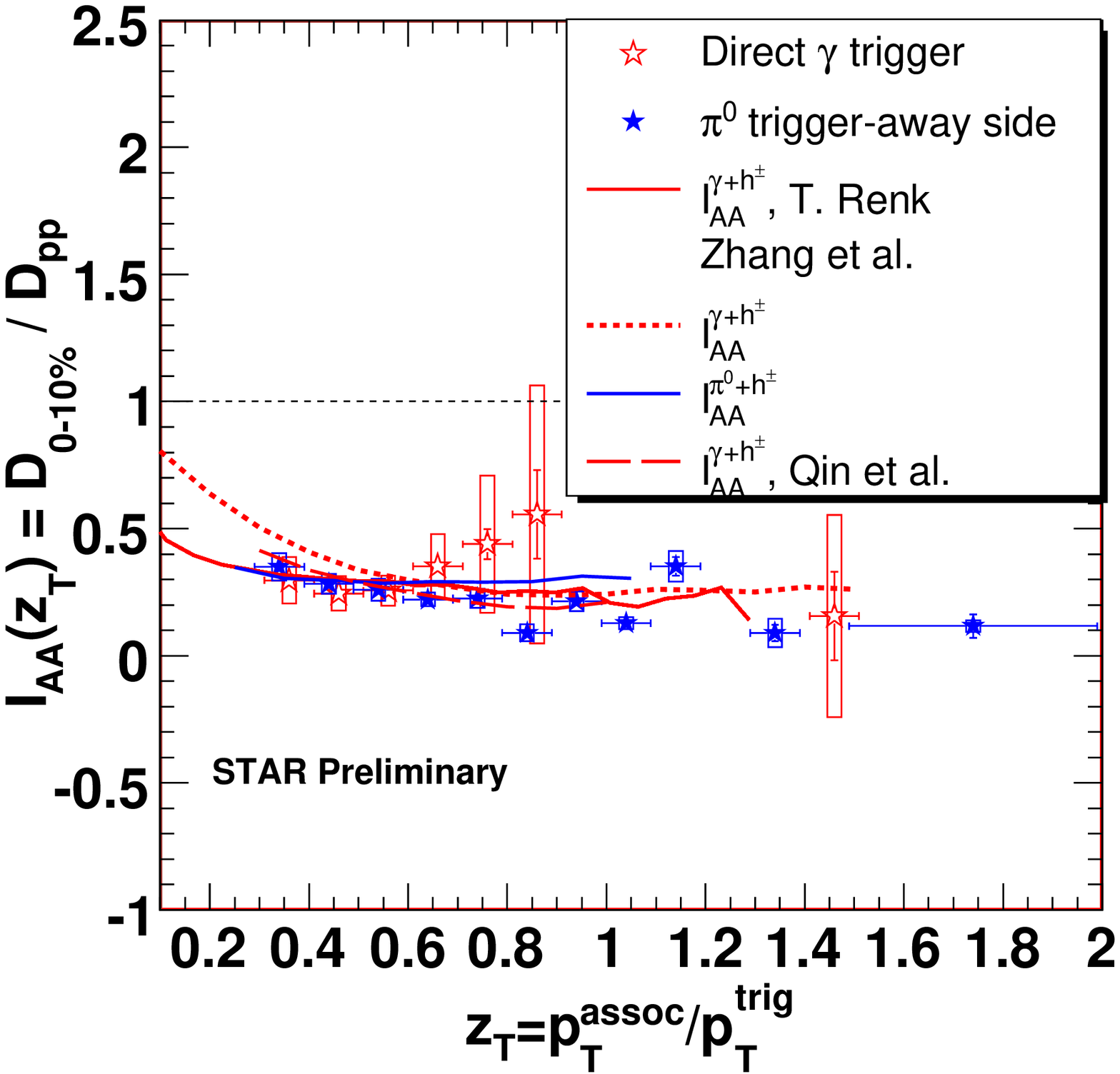}} 
\hspace{0.5in}
\resizebox{2in}{2in}{\includegraphics[3in,3in][7.5in,7.5in]{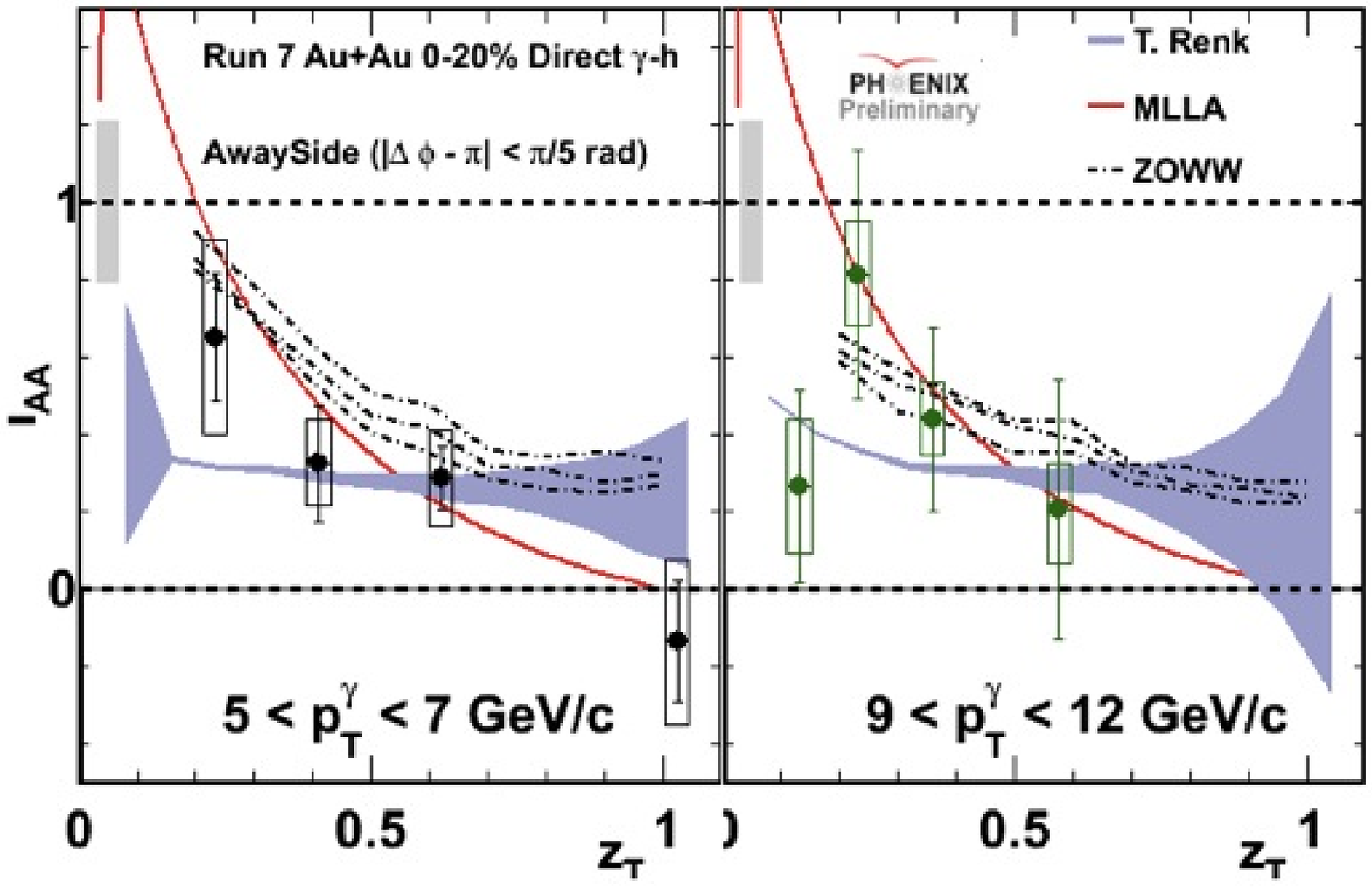}} 
    \caption{(Left plot) The $\gamma$ triggered and $\pi^0$ triggered associated yield on the away side from STAR. 
(Right plot) The $\gamma$ triggered associated yield from PHENIX. 
    Calculations  from the HT, ASW and AMY approaches have been superposed on both plots. In the right plot, the 
ASW calculations are represented by the dark gray bands.}
    \label{fig2}
\end{figure}

Once the constraint of single parameter fits is removed, the various pQCD-based energy loss schemes perform 
similarly and rather satisfactorily in explaining a wide variety of experimental data as demonstrated in Ref.~\cite{Renk:2009gn} 
for the ASW approach, in Ref.~\cite{wang} for the HT approach, as well as in Ref.~\cite{Qin:2009ff} for the AMY approach 
(see Ref.~\cite{jeon} for an explanation of the various acronyms). One of the major successes has been the 
parameter-free calculation of the $\gamma$  or hadron triggered away-side yield of leading hadrons. The experimental measurements from both STAR~\cite{hamed} and PHENIX~\cite{connors} and theoretical calculations for the $\gamma$ triggered yield are shown in Fig.~\ref{fig2}. While there are small differences at lower $p_T$  on the away side, the 
agreement between the various pQCD schemes at high $p_T$ (or $z_T$ as in the plots) is rather impressive. 

The calculation of the modification encountered by a hard parton in a given scheme depends, among other things, 
on the amount of transverse momentum imparted by the medium to the hard parton. 
This is encoded within the transport coefficient 
$\hat{q}$ which is defined as the transverse momentum squared per unit path length gained by the hard parton solely through collisions.
In numerical calculations, all pQCD-based approaches dial the value of $\hat{q}$ or some other parameter related to $\hat{q}$ 
to obtain agreement with one data point at one centrality. However there is wide disparity in this one value of $\hat{q}$ with 
the value from the $ASW$ scheme a factor of 5 larger than those from the HT and the AMY schemes.

This discrepancy remains unresolved and is a problem of some concern. The modification of hard jets in dense matter is sensitive to 
a limited number of properties of  the medium encoded as the standard transport coefficients. While the primary 
coefficient for light flavor energy loss is $\hat{q}$, the next two coefficients $\hat{e}$ (the rate of elastic energy loss 
per unit length) and $\hat{f}_{ab}$ [the rate of flavor rotation (from $a$ to $b$) per unit length] 
also play noticeable roles in heavy flavor 
loss and flavor dependence of the $R_{AA}$ respectively. The large uncertainty in $\hat{q}$ represents a major 
impediment in the use of jet modification to extract properties of the medium. The resolution of this problem is now 
one of the stated goals of the Theory and Experimental Collaboration for Hot Qcd Matter (TECHQM)~\cite{techqm}.

\section{Challenges to pQCD: Ridge, Cone and heavy flavor loss}

Beyond the single inclusive suppression and correlations at high $p_{T}$ are the unique structures seen in the 
high $p_{T}$ triggered yield of lower $p_{T}$ hadrons: A long ``ridge'' like structure is seen in the near side polar 
angle distribution and a conical structure in the away side azimuthal  distribution. These, along with the large  
suppression in heavy-flavor, have remained difficult problems for any pQCD-based approach to both formulate and 
quantitatively justify.

At the time of writing, the rather extensive data on the ridge~\cite{netrakanti}, particularly on the possible existence of the ridge at 
4 units of rapidity away from the trigger (along with the existence of a similar structure in untriggered two-particle 
correlations), have cast doubt on the likelihood of this being due to jet correlations. 
A set of non-final state models such as \emph{glasma flux tubes}~\cite{lappi} and 
flow of initial state radiation~\cite{shuryak} have been proposed 
as possible explanations of this phenomenon. As pointed out in Ref.~\cite{nagle}, models which have attempted 
to explain the ridge as a rapidity broadening of the final state radiation of the hard jet produce a more Gaussian 
distribution in rapidity with the signal depleting beyond 2 units of rapidity from the trigger.  

In spite of these issues on the near side, the only explanation for the conical structure on the away side 
so far has been due to the passage of jets through the dense matter. The leading explanation is 
believed to be due to the hydrodynamic response of the medium, in the form of a Mach shock, to the energy 
deposited by a hard jet. 
At this meeting, both STAR~\cite{netrakanti,konzer} and PHENIX~\cite{sickles,holzmann} have presented 
precise data on the dependence of the cone angle on the angle of the trigger to the reaction plane with the 
STAR data seeming to indicate the possibility of a turn-off of the conical pattern for in-plane emission. 

On the theory side, a 
complete explanation of this phenomenon requires two steps: The incorporation of the energy deposited 
by a hard jet within the same formalism as the energy loss by the hard parton, to understand the space-time 
profile of the energy momentum deposition. Secondly, a calculation of how the energy momentum tensor of the medium, predominantly the 
soft modes responsible for hydrodynamic behavior, is modified due the energy momentum deposited by the jet. 
At this meeting a solution to the first of these problems was presented. In Ref.~\cite{neufeld},
the authors realized that elastic energy loss by a single gluon is the energy deposited in the medium, 
the total amount of which depends on the number of radiated gluons in the in-medium shower (see Ref.~\cite{Qin:2009uh} for 
a more complete solution to this problem).

On the heavy flavor side, the difficulty of pQCD-based approaches to quantitatively describe the observed suppression continued 
at this meeting. Various new ideas have been proposed, such as the possible effect of heavy meson formation and 
dissociation in the QGP~\cite{sharma}, along with continuing developments in a more rigorous formulation of  the 
elastic and radiative energy lost by a massive quark in a medium with dynamical scattering centers~\cite{djordjevic}.
Both of these avenues are still in a development phase, however, the tentative results are promising.

 Alternative ideas beyond pQCD seem to have more success with this problem, in particular the 
approach of Ref.~\cite{rapp}, which models the cross section of bare heavy quarks travelling within the plasma 
via resonance scattering off anti-quarks in the medium. Yet another approach, which has enjoyed some success, 
also based on non-perturbative ideas, 
is that of AdS/CFT (see subsequent section). 

On the experimental side, progress has been two-fold: new corrections from $J/\psi$  decay in the cocktail 
have led to an approximate 15\% upward shift of the $R_{AA}$ of non-photonic electrons~\cite{dion}. 
Also heavy flavor triggered yields of associated hadrons as a function of azimuthal angle were presented 
for the first time at this meeting~\cite{engelmore_biritz}. While the error bars in the data are still too large to make even qualitative 
conclusions, this represents an extremely important measurement as it imposes a severe test of the Mach 
cone explanation of the away side azimuthal structure. Since heavy quarks travel at a slower 
speed than light quarks, the Mach angle is expected to shrink for heavy flavor triggers.

\section{Alternative approaches: AdS/CFT correspondence}

Despite the success of the approaches based on pQCD in explaining a wide variety of data 
on light flavor suppression and correlations, there have remained two major challenges to 
pQCD-based approaches: a description of the magnitude of heavy flavor suppression and 
a formulation of the space-time distribution of the energy deposition and the medium response. 
Both of these issues have been successfully combined into a single formalism within the AdS/CFT 
approach, where instead of solving the problem in QCD one attempts a solution in $\mathcal{N} = 4$ 
super symmetric Yang-Mills theory at strong coupling by calculating in the dual supergravity theory in AdS$_{5} \times S_{5}$.

In one version of this approach one describes the energy loss and deposition in a medium by a 
very heavy quark as a string trailing from a D7 brane on the 4-D boundary into the $5^{th} $ dimension of AdS$_{5}$ down 
to the black hole which corresponds to the finite temperature environment. New developments reported 
at this meeting consisted of extensions of this formalism to understand light flavor energy loss~\cite{gubser}. At present this 
has only been done for a gluon where one obtains a cubic length dependence of the energy loss i.e., 
$\Delta E  \sim L^{3}$. Thus one obtains a much stronger light flavor energy loss. There have also 
been the first phenomenological calculations based on AdS/CFT heavy flavor energy loss which show 
good agreement with the data on non-photonic $e^{-}$ $R_{AA}$~\cite{akamatsu}. 
The hadronization of the surviving portion of the hard jet remains a challenge in this approach and hence there exist no 
direct calculations of the $R_{AA}$.

On the energy deposition front, new developments have been presented in Ref.~\cite{chesler} which 
demonstrate a rising energy deposition rate (this is also the case in the pQCD calculation)
terminating in a possible Bragg peak. In Ref.~\cite{betz}, it has been argued that the strong 
cone pattern from AdS/CFT is necessary to ensure the survival of the correlation post freeze-out. 
It should be pointed out that the authors have made this claim based on an isochronous 
freeze-out of a static medium and assuming a constant rate of energy deposition for pQCD quenching. 
Other novel results presented at this 
meeting included a first calculation of $\hat{q}$ in a gravity dual of a non-conformal gauge theory~\cite{mia} and further 
developments in heavy quark drag in different backgrounds~\cite{horowitz}.

\section{The new waves: Event shapes and Monte-Carlo implementations}

From the start of the program at RHIC, jet quenching has always been limited to the study of the suppression 
of the leading hadron or the correlation between the leading and at most two other hadrons in the same or 
away side jet. This changed at the current meeting where the first data on full jet reconstruction in 
both $p$-$p$ and in $A$-$A$ were presented~\cite{salur,ploskon,lai}.

\begin{figure}[!htb]
\hspace{0.5in}
\resizebox{1.75in}{2in}{\includegraphics[1.5in,0.3in][6.7in,5.5in]{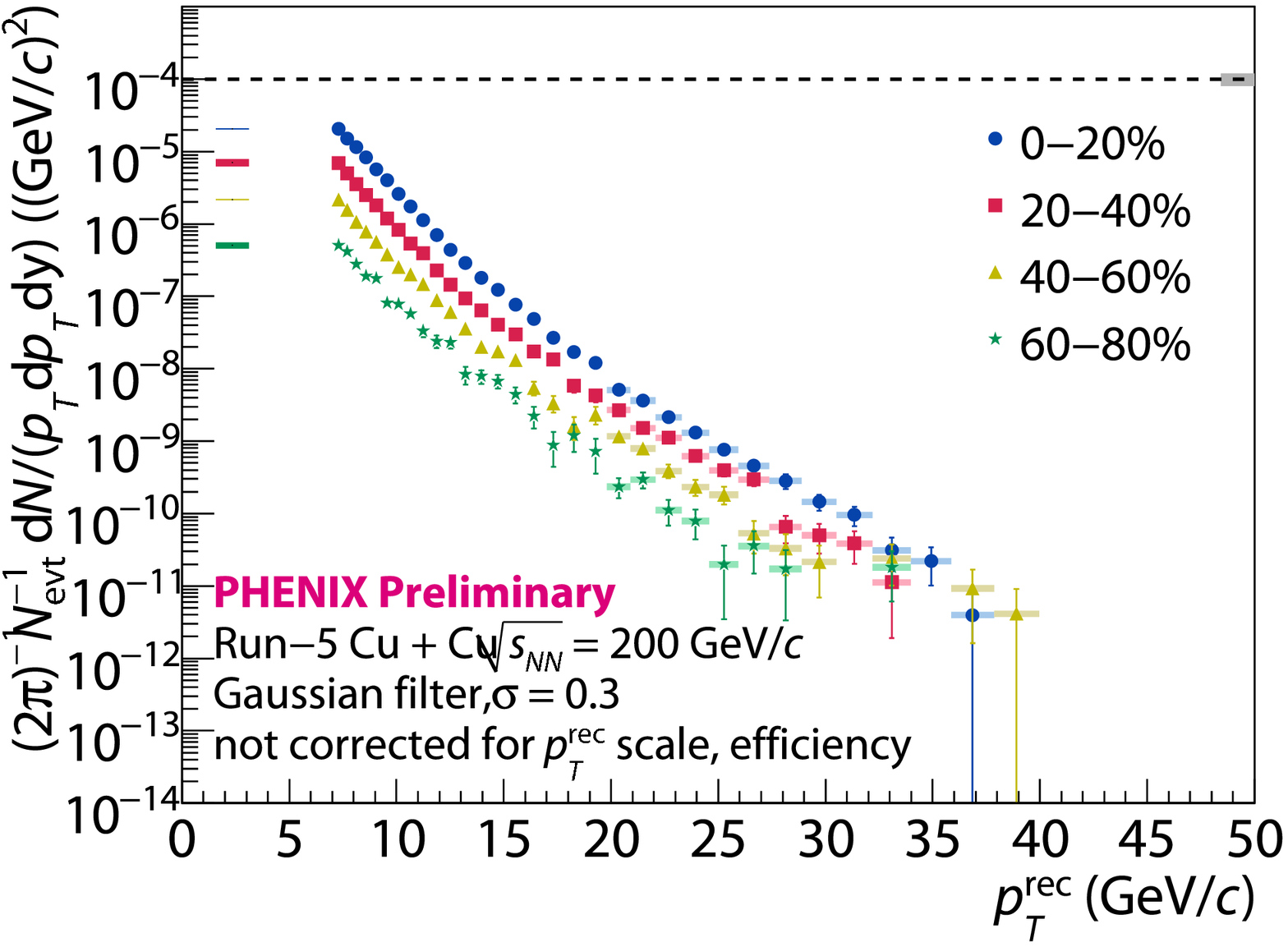}} 
\hspace{0.5in}
\resizebox{2in}{2in}{\includegraphics[0.5in,0.5in][6in,6.5in]{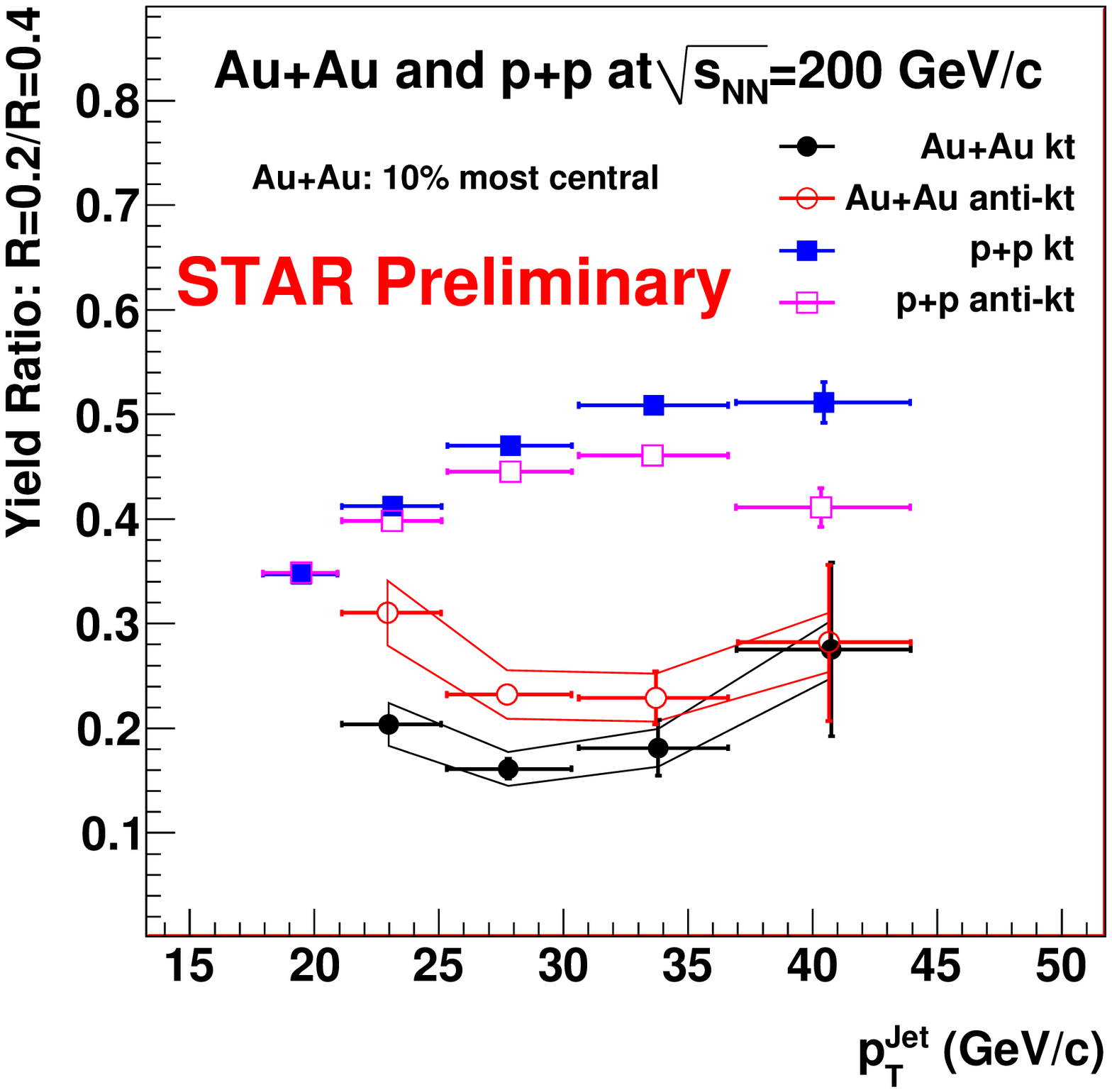}} 
    \caption{(Left plot) The full jet cross section in $Cu$-$Cu$ as reported by PHENIX. 
(Right plot) The ratio of yields in 
    two cone sizes in $p$-$p$ and $Au$-$Au$ as reported by STAR.  }
    \label{fig3}
\end{figure}

There remain differences in the jet definitions used by STAR and PHENIX: STAR uses the standard $k_{T}$ and 
anti-$k_{T}$  algorithms whereas PHENIX uses a Gaussian filter. 
However, results from both collaborations are rather 
spectacular. PHENIX has been able to reconstruct jets in $Cu$-$Cu$ up to 40 GeV~\cite{lai} (see Fig.~\ref{fig3}), whereas STAR has been 
able to reconstruct jets in both $p$-$p$ and $Au$-$Au$ up to 50 GeV for two different cone radii and also 
presented the jet $R_{AA}$.  The analysis of the drop in jet $R_{AA}$ for smaller cone radii clearly 
demonstrates that jets are broadened in the dense environment of a heavy-ion collision. Full jet reconstruction
may also allow for a first measurement of the medium modified fragmentation functions~\cite{caines_bruna} and 
dihadron fragmentation functions~\cite{elnimr}.

Rigorous theory calculations of full jet variables is still a few years in the future, though 
early attempts were presented at this meeting~\cite{zhang}. A complete analytical calculation of 
jet shapes requires a derivation of the ordering of both soft and hard induced radiation in the medium. 
To date most calculations are still performed in the soft gluon regime and rigorous calculations are limited 
to only the single gluon emission kernel. 

For a number of approaches, such extensions turn out to be analytically prohibitive and these have 
resorted to Monte-Carlo techniques. At this meeting four different developing Monte-Carlo schemes were 
introduced. Each of these uses a set of simple probabilistic techniques to mimic the LPM effect in the 
induced radiation. Implementations such as Q-PYTHIA~\cite{salgado} motivate a Sudakov form factor 
using the pre-calculated single gluon radiation probability in ASW and thus construct a probabilistic implementation. In 
JEWEL~\cite{zapp}, the random process of scattering is used to modify the formation time of the radiated gluon in a stochastic 
process. In YaJEM~\cite{Renk:2009gn}, the virtuality of the intermediate state is modified due to scattering in the medium leading to a change 
in the radiation pattern. In MARTINI~\cite{jeon}, analytical energy loss routines are used to modify the shower distribution in PYTHIA 
before hadronization. 
Monte-Carlo techniques, when properly developed, represent a promising approach in the 
calculation of the in-medium modification of complex observables that will occur in full jet reconstruction. 

\section{Conclusions}

QM 2009, was the last Quark Matter meeting prior to the turn on of the LHC heavy-ion program. In the nine 
years of RHIC running, spectacular progress has been made in the area of jet-medium interactions (a field non-existent at the SPS). 
There has been considerable success of the pQCD-based approaches as well as challenges to this paradigm. 
This has led to a considerable amount of effort, both theoretical and phenomenological, to extend pQCD-based descriptions of 
jet-medium interactions.
In the last few years, spurred by these challenges, the entirely new field of strong coupling energy loss has appeared. 

Experimental data in the 
field are quite extensive and are now statistically significant enough to become discriminatory, even between closely related 
schemes. At this meeting, entirely new data on full jet reconstruction were presented. The complete explanation 
of the various observables will require new extensions in theory. This will in turn lead to the development of 
jet-medium interactions as more comprehensive probes of the dense deconfined matter produced at RHIC and the LHC. 

The author acknowledges the hospitality of the Physics Department at Duke University where these proceedings were 
written. These proceedings and the presentation are the outcome of extensive discussions with (and in some cases tutoring by) 
various speakers and participants at Quark Matter whom the author wishes to thank. Special mention is made of 
Meghan~Connors, Ahmed~Hamed, Mateusz~Ploskon and Rui~Wei who modified their plots exclusively 
for these proceedings and the presentation.
The author also thanks Ulrich Heinz, Berndt M\"{u}ller 
and Guang-You Qin for discussions and a careful reading of the manuscript.
This work was supported in part by U.S. 
Department of Energy under grant No. DE-FG02-01ER41190.

\end{document}